\documentclass[prd,nofootinbib,english]{revtex4}

\usepackage{graphicx,float}
\usepackage{amsmath,amssymb,amsfonts}
\usepackage{epsfig,color}
\usepackage[thinlines]{easytable}
\usepackage{pdfpages}
\usepackage{array}
\usepackage{cancel}
\usepackage{mathtools}
\usepackage{accents}
\usepackage{subfigure}
\usepackage{enumitem}
\usepackage[dvipsnames]{xcolor}
\usepackage{refstyle}
 \usepackage{hyperref}
\hypersetup{
    colorlinks=true,
    linkcolor=blue,
    filecolor=magenta,      
   citecolor=blue
}

\def\be{\begin{equation}}
	\def\ee{\end{equation}}
\def\bea{\begin{eqnarray}}
	\def\eea{\end{eqnarray}}

\begin{document}

\title{A simple parity violating model in the symmetric teleparallel gravity and its cosmological perturbations}

\author{Mingzhe Li}
\email{limz@ustc.edu.cn}
\author{Dehao Zhao}
\email{dhzhao@mail.ustc.edu.cn}
\affiliation{Interdisciplinary Center for Theoretical Study,
	University of Science and Technology of China, Hefei, Anhui 230026, China}
\affiliation{Peng Huanwu Center for Fundamental Theory, Hefei, Anhui 230026, China}

\begin{abstract}
	Parity violating models based on the symmetric teleparallel gravity have been considered in the literature, but their applications in cosmology and especially the features in cosmological perturbations have not been fully explored. In this paper we consider such a model which modifies general relativity merely through a parity non-conserved coupling, $\phi(x) Q\widetilde{Q}$, within the framework of the symmetric teleparallel gravity. We study in detail its cosmological applications, especially its linear perturbation theory. Besides the already known parity violation in the tensor perturbations, we find that the vector perturbations in this model are promoted to be dynamical degrees of freedom, and the left- and right-handed vector modes propagate with different velocities. More importantly, one of the vector modes is found to be a ghost at high momentum scales, this will give rise to the vacuum instability in the quantum theory of cosmological perturbations. 
\end{abstract}
\maketitle

\section{Introduction}

Discrete symmetries, including parity, charge conjugation and time reversal, have played important roles in discovering fundamental physical laws. Since parity was found to be broken in weak interactions \cite{Lee:1956qn}, we know that most of them and their combinations are not exact in nature. An interesting question one may ask is whether or not these discrete symmetries are broken in the gravity sector. On the other hand, although Einstein's General Relativity (GR) has achieved great success, it is often considered as an effective theory and thought to be modified at high energy and/or large length scales. Whether the discrete symmetries are preserved in these modifications or extensions is an unknown problem and expected to be tested by experiments in the future.  Recently there were lots of theoretical studies on the parity violating gravities in the literature. These were stimulated by the experimental discoveries of gravitational waves (GWs) \cite{Abbott:2016blz} and the developments in the cosmic microwave background radiation (CMB) experiments \cite{Li:2017drr}, which are planned to find primordial GWs originated from the early universe. If there are parity violations in gravity, the left- and right-handed polarized GWs will propagate differently and corresponding signals are expected to be captured by well-designed experiments. 

A famous parity violating gravity model is the Chern-Simons (CS) modified gravity \cite{Jackiw:2003pm,Alexander:2009tp}, which modifies GR by a CS Lagrangian density: $\mathcal{L}_{CS}\sim  \phi(x)R\widetilde{R}$, here $R\widetilde{R}\equiv \epsilon^{\mu\nu\rho\sigma} {\mathring{R}_{\mu\nu}}{}^{\alpha\beta}\mathring{R}_{\rho\sigma\alpha\beta}$ is the gravitational Pontryagin density, $\mathring{R}_{\alpha\beta\rho\sigma}$ is the Riemann tensor constructed from the Levi-Civita connection, and $\epsilon^{\mu\nu\rho\sigma}$ is the four dimensional Levi-Civita tensor relating to the antisymmetric symbol $e^{\mu\nu\rho\sigma}$ and the determinant $g$ of the metric as $\epsilon^{\mu\nu\rho\sigma}=e^{\mu\nu\rho\sigma}/\sqrt{-g}$. The coupling scalar field $\phi(x)$ is indispensable, without it the CS Lagrangian will become a surface term. In the linear perturbation theory, the CS modified gravity indeed makes a difference between the left- and right- handed polarizations of GWs, but suffers from the ghost problem. As pointed out in Ref. \cite{Dyda:2012rj}, when going to high momentum scales one of the polarized components of GWs has a wrong sign in its kinetic term in the Lagrangian. This corresponds to the ghost mode. The existence of the ghost mode indicates pathological behaviors in the model, for example there is no stable vacuum state in the quantum theory. Further extensions to the CS modified gravity were made in Refs. \cite{Crisostomi:2017ugk, Gao:2019liu} where more parity violating interactions were introduced into the Lagrangian. Unfortunately these did not prevent the appearance of the ghost mode at high frequencies, as shown explicitly by Eqs.(3.5)-(3.6) and (3.39)-(3.40) in Ref.\cite{Bartolo:2020gsh}.

Within the framework of Riemannian geometry, the CS term is the simplest one we can construct to produce parity violation. So if we want to have parity violating gravity models which not only have forms as simple as that of the CS modified gravity but also can avoid the ghost mode, we must go beyond the Riemannian geometry. Recently, such a kind of parity violating gravity model \cite{Li:2020xjt,Li:2021wij} was proposed within the framework of teleparallel gravity (TG) (see Ref. \cite {Bahamonde:2021gfp} for a recent review). The TG theory can be considered as a constrained metric-affine theory. It is formulated in the spacetime endowed with zero curvature and metric compatible connection. The gravity is identified with the spacetime torsion. The TG model we are most concerned about is the so-called Teleparallel Equivalent of General Relativity (TEGR), which is equivalent to GR and its action is equal to the Einstein-Hilbert action up to a surface term. The parity violating gravity model in Refs. \cite{Li:2020xjt,Li:2021wij} modifies TEGR by a Nieh-Yan (NY) Lagrangian density: $\mathcal{L}_{NY}  \sim  \phi(x) T\widetilde{T}$, here $T\widetilde{T}\equiv\epsilon^{\mu\nu\rho\sigma}\eta_{AB}{T^A}_{\mu\nu}{T^{B}}_{\rho\sigma}$, $\eta_{AB}$ is the Minkowski metric in the local space, and ${T^{A}}_{\rho\sigma}$ is the torsion two form\footnote{In general the NY density contains another term proportional to the curvature tensor, but within the framework of TG the curvature vanishes and the NY density reduces to the one discussed here. Similar coupling of NY density to a non-dynamical $\theta$ parameter was considered in Ref. \cite{Chatzistavrakidis:2020wum}.}. We may call this model as Nieh-Yan modified teleparallel gravity (NYTG). Some properties, especially the cosmological perturbations (including all the scalar, vector and tensor perturbations) of this model have been investigated in detail in Refs. \cite{Li:2020xjt,Li:2021wij}. It was found that besides the parity violating signal in GWs, the NY coupling has important effects on the scalar perturbations, especially in the spatially flat universe it eliminates one scalar degree of freedom at the linear perturbation level, due to the existence of an extra constraint. But the vector perturbations remain the same as in GR. More importantly, the NYTG model was found to be free from ghost, at least up to the linear order. Besides the cosmological dynamics, the parametrized post-Newtonian limit of NYTG was recently explored in Ref. \cite{Rao:2021azn}. In addition, more parity violating interactions than the NY term was considered in Ref. \cite{Hohmann:2020dgy}.

Besides TG, there is another similar non-Riemannian framework to build gravity theories, that is the so-called Symmetric Teleparallel Gravity (STG)  \cite{Nester:1998mp}. The STG theory is formulated in the spacetime endowed with zero curvature and zero torsion, and attributes the gravity to the non-metricity.  There is also a STG model equivalent to GR, i.e., the Symmetric Teleparallel General Relativity (STGR), in which the Lagrangian density is a quadratic polynomial of the non-metricity tensor and is equal to the Einstein-Hilbert Lagrangian density up to a surface term. Similarly, within this framework one can modify STGR by a parity violating interaction:  $\mathcal{L}_{PV}  \sim  \phi(x) Q\widetilde{Q}$, here $Q\tilde{Q}\equiv\epsilon^{\mu\nu\rho\sigma} {Q}_{\mu\nu\alpha}{{Q}_{\rho\sigma}}^{\alpha}$ with ${Q}_{\mu\nu\alpha}$ the non-metricity tensor. This is very similar to $\phi(x)R\widetilde{R}$ in the CS modified gravity and $\phi(x)T\widetilde{T}$ in the NYTG model. Nevertheless there is a significant difference: the densities $R\widetilde{R}$ and $T\widetilde{T}$ in the latter two models are topological terms, but $Q\widetilde{Q}$ mentioned here is not. So the scalar field $\phi(x)$ is not necessarily a dynamical field, it can be a constant. 
In fact the coupling $\phi(x) Q\widetilde{Q}$ together with other modifications to STGR have already been considered in Ref. \cite{Conroy:2019ibo}, where its effects on the experimental signals of GWs were studied. To our knowledge, its properties, especially its consequences on cosmology have not been fully explored. In this paper we will consider the simple model in which the parity violating term $\phi(x) Q\widetilde{Q}$ is the only modification to STGR. We will investigate in detail the changes by this model to the background evolution of the universe and to the scalar, vector and tensor perturbations. We find surprisingly that the vector perturbations in this model are dynamical components and one of them must be a ghost mode at high momentum scales. 

This paper is organized as follows. In section \ref{STGR introduction} we will introduce this simple parity violating gravity model. In section \ref{cosmology application}, we will consider its cosmological applications, and present our main results about the studies on the cosmological perturbations. Section \ref{summary} is the summary.

\section{Parity violating model based on the symmetric teleparallel gravity} \label{STGR introduction}

From now on, we will use the unit $8\pi G=1$ and the convention of most negative signature for the metric. The Greeks $\mu,\nu,\sigma,...=0,1,2,3$ are used to represent spacetime tensor indices, and the Latins $i,j,k,...=1,2,3$ denote the spatial components. The STG theory can be considered as a constrained metric-affine theory, in which one starts from a spacetime with a metric $g_{\mu\nu}$ and an affine connection ${\Gamma^{\lambda}}_{\mu\nu}$ constrained by the zero-curvature and zero-torsion conditions:
\bea\label{noncurvature}
{R^{\rho}}_{\sigma\mu\nu}\equiv \partial_{\mu}{\Gamma^{\rho}}_{\nu\sigma}+{\Gamma^{\rho}}_{\mu\alpha}{\Gamma^{\alpha}}_{\nu\sigma}-\{\mu\leftrightarrow \nu\}=0~,  {T^{\rho}}_{\mu\nu}\equiv {\Gamma^{\rho}}_{\mu\nu}-{\Gamma^{\rho}}_{\nu\mu}=0~.
\eea
The gravity in this theory is identified with the non-metricity. As usual the non-metricity tensor is defined as
\begin{eqnarray}\label{qdefinition}
Q_{\alpha \mu \nu}\equiv \nabla_{\alpha} g_{\mu \nu} = \partial_{\alpha} g_{\mu\nu} - {\Gamma^{\lambda}}_{\alpha\mu}g_{\lambda\nu} - {\Gamma^{\lambda}}_{\alpha\nu}g_{\mu\lambda}~,
\end{eqnarray}
which measures the failure of the affine connection to be metric-compatible. The STGR model has the following action,
\begin{align}\label{action Q}
S_{g}&=\frac{1}{2} \int d^{4}x \sqrt{-g}\mathbb{Q}\nonumber\\
&\equiv \frac{1}{2} \int d^{4}x \; \sqrt{-g} [  \frac{1}{4}Q_{\alpha\mu\nu}Q^{\alpha\mu\nu} -\frac{1}{2}Q_{\alpha\mu\nu}Q^{\mu\nu\alpha}-\frac{1}{4}Q^{\alpha}Q_{\alpha}+\frac{1}{2}{\bar{Q}}^{\alpha}Q_{\alpha}   ]  ,
\end{align}
where $\mathbb{Q}$ is the non-metricity scalar and the vectors $Q_{\alpha}, \bar{Q}_{\alpha}$ are two different traces of the non-metricity tensor, $Q_{\alpha} = g^{\sigma\lambda}Q_{\alpha\sigma\lambda}, \quad  \bar{Q}_{\alpha} =  g^{\sigma\lambda}Q_{\sigma\alpha\lambda}$. 
In terms of the zero-curvature and zero-torsion constraints (\ref{noncurvature}) and the general relation between the curvature from ${\Gamma^{\rho}}_{\mu\nu}$ and the curvature from the Levi-Civita connection:
\begin{eqnarray} \label{curvature relation}
{{R^{\sigma }}_{\rho\mu\nu}}  = {{\mathring{R}^{\sigma }}_{\rho\mu\nu}}  + 2 \mathring{\nabla}_{[\mu}{S^{\sigma}}_{\nu]\rho}-2{S^{\lambda}}_{[\mu|\rho|}{S^{\sigma}}_{\nu]\lambda}~,
\end{eqnarray}
where the distortion tensor is defined as
\begin{eqnarray}
S_{\lambda\mu\nu}= -\frac{1}{2}( T_{\mu\nu\lambda} +T_{\nu\mu\lambda} - T_{\lambda\mu\nu} )
-\frac{1}{2}( Q_{\mu\nu\lambda} +Q_{\nu\mu\lambda} - Q_{\lambda\mu\nu} )~,
\end{eqnarray}
one can easily find that the action (\ref{action Q}) is equal to the Einstein-Hilbert action up to a boundary term,
\begin{equation}\label{action-g2}
S_g= \frac{1}{2}\int d^{4}x   \sqrt{-g}\left[  -\mathring{R}-\mathring{\nabla}_{\mu}(Q^{\mu} - \bar{Q}^{\mu}) \right] ~. 
\end{equation}
In above we have used $\mathring{\nabla}$ to denote the covariant derivative associated with the Levi-Civita connection.  

Here we would like to comment on the fundamental variables of STG theory. 
With the zero-curvature condition one can represent the affine connection generally as
\begin{eqnarray}\label{Omega}
{\Gamma^{\lambda}}_{\mu \nu}  = {(\Omega^{-1})^{\lambda}}_{\sigma} \partial_{\mu} {\Omega^{\sigma}}_{\nu}~,
\end{eqnarray}
where ${\Omega^{\alpha}}_{\beta}$ is an arbitrary element of the group GL(4) and has non-zero determinant. Furthermore, the zero-torsion condition requires that ${\Omega^{\alpha}}_{\beta}$ should be expressed as ${\Omega^{\alpha}}_{\beta}=\partial y^{\alpha}/ \partial x^{\beta}$, here $y^{\alpha}(x)$ are four functions and totally determine all the components the affine connection, 
\begin{eqnarray} \label{gamma=}
{\Gamma^{\lambda}}_{\mu \nu} (x^{\mu}) = \frac{\partial x^{\lambda}}{\partial y^{\beta}} \partial_{\mu}\partial_{\nu} y^{\beta}~.
\end{eqnarray} 
So, in a STG theory one can take the metric $g_{\mu\nu}$ and the functions $y^{\alpha}(x)$ as independent variables. 
From Eq. (\ref{gamma=}), one can see that the four functions $y^{\alpha}$ constitute a special coordinate system in which the affine connection vanishes, i.e. ${\Gamma^{\lambda}}_{\mu \nu} (y^{\alpha})= 0$. Mathematically, with the zero-curvature and zero-torsion constraints one can always obtain a special coordinate system with zero affine connection \cite{Bao:2000}. One can fix to this special coordinate system for studies. Once taking this gauge, only the ten components of the metric are considered as fundamental variables. In fact this ``coincident gauge" has been extensively used in studies on STG theories to simplify the calculations. 
However, taking the coincident gauge at the beginning usually breaks the diffeomorphism invariance explicitly. A unique exception is the STGR model (\ref{action Q}) in which the affine connection vanishes or not merely bings changes to the boundary term, as can be seen from Eq. (\ref{action-g2}). In general cases, one can still take this gauge even though the diffeomorphism invariance is broken explicitly. But sometimes in these cases we should be careful that the special coordinate system $y^{\alpha}$ and some frequently used conventional parametrizations for the metric are not compatible \cite{Zhao:2021zab}. In this paper, except in the subsection \ref{discussion}, we will not take the coincident gauge and will treat the ten components of the metric together with the four functions $y^{\alpha}$ as fundamental variables.

There are many researches on modified STGR in the literature, for instances the $f(\mathbb{Q})$ models \cite{Lu:2019hra,Lazkoz:2019sjl} in which the non-metricity scalar in the action (\ref{action Q}) is replaced by its arbitrary functions, and the ``Newer GR" \cite{Runkla:2018xrv,Soudi:2018dhv,Hohmann:2018wxu} in which the action is constructed by arbitrary combinations of the terms which are parity even and quadratic in the non-metricity tensor. The parity violating quadratic term $\phi(x) Q\widetilde{Q}$ has been considered in Ref. \cite{Conroy:2019ibo}, and recently in Ref. \cite{Bombacigno:2021bpk} in a more general metric-affine case, but up to now its properties have not been fully explored. 

In this paper, we will reconsider the parity violating interaction $\phi(x) Q\widetilde{Q}$ as an extension to the STGR model (\ref{action Q}), this is equivalent to modifying GR slightly by a parity violating interaction. The full action of the model we consider is
\begin{align} \label{action full}
S&=\int d^{4}x \sqrt{-g} \left( \frac{\mathbb{Q}}{2} - c\phi \epsilon^{\mu\nu\rho\sigma}Q_{\mu\nu\alpha}{Q_{\rho\sigma}}^{\alpha} \right) +S_{\phi}+S_{m} ~,\nonumber\\
&=\int d^{4}x \sqrt{-g} \left(- \frac{\mathring{R}}{2} - c\phi \epsilon^{\mu\nu\rho\sigma}Q_{\mu\nu\alpha}{Q_{\rho\sigma}}^{\alpha} \right)  - \frac{1}{2}\int d^{4}x   \sqrt{-g}\left[  \mathring{\nabla}_{\mu}(Q^{\mu} - \bar{Q}^{\mu}) \right]+S_{\phi}+S_{m} ~,
\end{align} 
where $c$ in the parity violating term is a coupling constant, the non-metricity tensor $Q_{\mu\nu\alpha}$ is given by Eq. (\ref{qdefinition}) together with Eq. (\ref{gamma=}). We also take into account the action $S_{\phi}$ for the scalar field 
\begin{equation}
S_{\phi}=\int d^{4}x \sqrt{-g} \left[\frac{1}{2}g^{\mu\nu}\partial_{\mu}\phi\partial_{\nu}\phi-V(\phi)\right]~,
\end{equation}
and $S_{m}$ for other matter. The surface term at the third line of Eq. (\ref{action full}) can be dropped out since it has no effects on the equations of motion of the system. 
Besides the simple form of this modification, it does not contain any higher oder derivatives of the metric and excludes the potential ghost modes brought by higher derivatives. This is especially clear when adopting the coincident gauge. But we still need to check whether there are other type of pathologies left. 

As mentioned before, we will treat metric $g_{\mu\nu}$ and functions $y^{\mu}$ as independent and fundamental variables in the gravity sector. Through the variations of the action (\ref{action full}) with respect to the metric and the scalar field $\phi$, we obtain the following equations of motion (EOMs)
\begin{eqnarray}
& &	\mathring{G}^{\mu\nu} +N^{\mu\nu} = {T_{m}}^{\mu\nu} + {T_{\phi}}^{\mu\nu} \label{metric equation},\label{eom1}\\
& &	\mathring{\nabla}^{\mu}\mathring{\nabla}_{\mu}\phi +\frac{dV}{d\phi} +c\epsilon^{\mu\nu\rho\sigma}Q_{\mu\nu\alpha}{Q_{\rho\sigma}}^{\alpha}=0, \label{scalar EOM}
\end{eqnarray}
where $\mathring{G}^{\mu\nu}=\mathring{R}^{\mu\nu}-\mathring{R}g^{\mu\nu}/2$ is the Einstein tensor, ${T_{m}}^{\mu\nu}=(-2/\sqrt{-g})(\delta S_{m}/\delta g_{\mu\nu})$ and ${T_{\phi}}^{\mu\nu}=\mathring{\nabla}^{\mu}\phi \mathring{\nabla}^{\nu}\phi -\left(\frac{1}{2}g^{\alpha\beta}\partial_{\alpha}\phi\partial_{\beta}\phi-V(\phi)\right)g^{\mu\nu}$ are the energy-momentum tensors for the matter and scalar field respectively, and the tensor $N^{\mu\nu}$ is defined as
\begin{eqnarray}
N^{\mu\nu}=\frac{2c}{\sqrt{-g}} \left[ \nabla_{\lambda}\left(2\phi e^{\rho\sigma\lambda(\mu}{Q_{\rho\sigma}}^{\nu)} \right) + \phi e^{\alpha\beta \rho\sigma}{Q_{\alpha\beta}}^{\mu}{Q_{\rho\sigma}}^{\nu} \right]~.
\end{eqnarray}

The variation with respect to $y^{\mu}$ needs more tricks. First we should consider the variation of the affine connection. In terms of Eq. (\ref{Omega}),  ${\Gamma^{\lambda}}_{\mu \nu}  = {(\Omega^{-1})^{\lambda}}_{\sigma} \partial_{\mu} {\Omega^{\sigma}}_{\nu}$, and the relation $\delta  {(\Omega^{-1})^{\lambda}}_{\sigma} =- {(\Omega^{-1})^{\lambda}}_{\alpha} {(\Omega^{-1})^{\beta}}_{\sigma}\delta {\Omega^{\alpha}}_{\beta}$, the variation of the connection has the following tensorial expression,
\begin{eqnarray}
\delta{ \Gamma^{\lambda}}_{\mu\nu}=\nabla_{\mu}{X^{\lambda}}_{\nu}~,~{\rm with}~{X^{\lambda}}_{\nu}= {(\Omega^{-1})^{\lambda}}_{\sigma}\delta  {\Omega^{\sigma}}_{\nu}~.
\end{eqnarray}
Furthermore, because ${\Omega^{\alpha}}_{\beta}=\partial y^{\alpha}/ \partial x^{\beta} $, the infinitesimal tensor ${X^{\lambda}}_{\nu}$ can be rewritten as,
\begin{eqnarray}
{X^{\lambda}}_{\nu}=\frac{\partial x^{\lambda}}{\partial y^{\sigma}} \partial_{\nu} \delta y^{\sigma}
=\partial_{\nu}\left(\frac{\partial x^{\lambda}}{\partial y^{\sigma}}\delta y^{\sigma}\right)-\partial_{\nu} {(\Omega^{-1})^{\lambda}}_{\sigma}\delta y^{\sigma}=\nabla_{\nu} X^{\lambda}~,
\end{eqnarray}
where the infinitesimal vector $X^{\lambda}$ is defined as $X^{\lambda}\equiv (\partial x^{\lambda}/\partial y^{\sigma}) \delta y^{\sigma}$.
With these considerations, one can rewrite the variation of the affine connection under the zero-curvature and zero-torsion constraints as
\begin{eqnarray} \label{variation of connection}
\delta{ \Gamma^{\lambda}}_{\mu\nu}=\nabla_{\mu}\nabla_{\nu}X^{\lambda}~. 
\end{eqnarray}
This fact has also been pointed out in Ref. \cite{Hohmann:2021fpr}.
If the matter other than the scalar field couples to the metric minimally,  one may obtain the following equation of motion through the variations of the action with respect to $y^{\lambda}$,
\begin{eqnarray}
\nabla_{\nu}\nabla_{\mu}\left( \phi e^{\mu\alpha\rho\sigma}{Q_{\rho\sigma}}^{\nu}g_{\lambda\alpha}  \right)=0~.\label{eom2}
\end{eqnarray}
However, this equation is not independent if the equations (\ref{eom1}), (\ref{scalar EOM}) and the equation of motion for other matter (equivalent to the covariant conservation law of its energy-momentum tensor) are already known. It can be easily checked that when acting the derivative operator $\mathring{\nabla}_{\mu}$ to Eq. (\ref{eom1}), then using the Bianchi identity, the conservation law $\mathring{\nabla}_{\mu} {T_m}^{\mu\nu}=0$ and Eq. (\ref{scalar EOM}), one can obtain Eq. (\ref{eom2}). 
This non-independence reflects the diffeomorphism invariance of the model (\ref{action full}) if we do not impose any gauge condition on the affine connection so that it has the general form (\ref{gamma=}). 
The diffeomorphism invariance requires that the action is invariant under the infinitesimal coordinate transformation $x^{\mu}\rightarrow x^{\mu}+\xi^{\mu}$, i.e.,
\begin{eqnarray}
\delta_{\xi}S = \frac{\delta S}{\delta g}|_{\Gamma, \phi, \cdot\cdot\cdot} \delta_{\xi}g + \frac{\delta S}{\delta \phi}|_{g,\Gamma, \cdot\cdot\cdot}\delta_{\xi}\phi+\frac{\delta S}{\delta \Gamma}|_{g,\phi, \cdot\cdot\cdot} \delta_{\xi}\Gamma+\cdot\cdot\cdot
=0~.
\end{eqnarray}
So if the EOMs of the metric, the scalar $\phi$ and other matter are satisfied, we will have $\frac{\delta S}{\delta g}|_{\Gamma, \phi, \cdot\cdot\cdot}=0$, $\frac{\delta S}{\delta \phi}|_{g,\Gamma, \cdot\cdot\cdot}=0$ and so on, then $\frac{\delta S}{\delta \Gamma}|_{g,\phi, \cdot\cdot\cdot} \delta_{\xi}\Gamma=0$ is satisfied automatically. Furthermore, we know that under the coordinate transformation the affine connection transforms as
\begin{eqnarray}
\delta_{\xi} {\Gamma^{\lambda}}_{\mu\nu}= -\nabla_{\mu}\nabla_{\nu}\xi^{\lambda} ~,
\end{eqnarray}
so the variation of connection, Eq.(\ref{variation of connection}), will be the same as its change brought by the coordinate transformation as long as we replace $X^{\lambda}$ with $- \xi^{\lambda}$. This means $\frac{\delta S}{\delta \Gamma}|_{g,\phi, \cdot\cdot\cdot} \delta_{\xi}\Gamma=0$ from the diffeomorphism invariance is equivalent to $\frac{\delta S}{\delta \Gamma}|_{g,\phi, \cdot\cdot\cdot} \delta_{y}\Gamma=0$ which implies the EOMs of $y^{\mu}$. This analysis is not only applicable to the current model but also to all other STG models as long as we keep the diffeomorphism invariance. So for these models, if we have already the EOMs for the metric and for all matter, we do not need to consider the EOM for $y^{\mu}$ independently anymore. For the model (\ref{action full}) studied here, we need only consider Eqs. (\ref{eom1}), (\ref{scalar EOM}) and the conservation law $\mathring{\nabla}_{\mu} {T_m}^{\mu\nu}=0$ as independent EOMs. These equations together with the action (\ref{action full}) are the bases for exploring the cosmological applications of this model in the rest of this paper.

\section{Cosmological Applications} \label{cosmology application}

From now on, we apply the parity violating gravity model (\ref{action full}) to cosmology. We first consider the background evolution of the universe and then attach importance to the cosmological perturbations. 

\subsection{Background Evolution} 

For simplicity, we take the spatially flat Friedmann-Robertson-Walker (FRW) universe as the background.  As usual the solution of the metric is contained in the line element which is assumed to have the simple form,
\begin{eqnarray}\label{FRW}
ds^2=a^{2}(\eta)(d\eta^{2}-\delta_{ij}dx^idx^j)~.
\end{eqnarray}
At the same time, the solution of the affine connection (not the Levi-Civita connection) can be assumed to be zero, ${\Gamma^{\lambda}}_{\mu\nu}=0$. This may be understood as taking the coincident gauge, but here it is understood as a part of solution to the whole system which is diffeomorphism invariant. These two different viewpoints will not cause problems. More importantly, these assumptions about the metric and the affine connections are reasonable for the spatially flat FRW universe.  According to the method by Ref. \cite{Zhao:2021zab}, it is easy to check that in this model these two assumptions are compatible. This means for the spatially flat FRW universe we can take the coordinate system $y^{\mu}=x^{\mu}$ in which the affine connection vanishes and simultaneously the metric is allowed to have the usual ansatz (\ref{FRW}). In addition, in the background the scalar field is homogeneous and other matter species are considered as homogeneously distributed fluids. Then the EOMs (\ref{eom1}) and (\ref{scalar EOM}) give the following equations, 
\begin{eqnarray}\label{beom}
3 \mathcal{H}^{2}=a^2(\rho_{\phi}+\rho_{m}), \quad -2\mathcal{H}'-\mathcal{H}^{2}=a^2(p_{\phi}+p_{m}),\quad \phi''+2\mathcal{H}\phi'+a^2V_{\phi}=0~,
\end{eqnarray}
where the prime represents the derivative with respect to the conformal time $\eta$, and $\mathcal{H}=a'/a$ is the conformal Hubble rate. As usual $\rho_{\phi}={\phi'}^2/(2a^2)+V$ and $p_{\phi}={\phi'}^2/(2a^2)-V$ are the energy density and pressure of the scalar field respectively. These equations are the same as those in GR, so we conclude that the parity violating coupling $\phi(x)Q\widetilde{Q}$ has no effect on the spatially-flat FRW background.

\subsection{Cosmological Perturbation Equations}

Now we turn to the linear perturbations around this background. With the Scalar-Vector-Tensor decomposition, the line element for the perturbed universe is generally written as 
\begin{align}\label{decomposition}
ds^2=a^2\{(1+2A)d\eta^2+2(\partial_{i}B+B_{i})d\eta dx^{i} -[(1-2\psi)\delta_{ij}+2\partial_{i}\partial_{j}E+\partial_{i}E_{j}+\partial_{j}E_{i}+h_{ij}]dx^{i}dx^{j}\}~,
\end{align}
where the vector perturbations $B_{i}$, $E_{i}$ are transverse, i.e., $\partial_{i}B_{i}=\partial_{i}E_{i}=0$, and the tensor perturbations, i.e., GWs $h_{ij}$ satisfy the transverse and traceless conditions, $\partial_{i}h_{ij}=\delta^{ij}h_{ij}=0$.
Since the affine connection vanishes in the background, itself is considered as small perturbation in the perturbed universe. In another word, in the background the four functions $y^{\mu}$ are the same with the coordinates $x^{\mu}$, but they do not match when perturbations are involved, i.e., $y^{\mu}=x^{\mu}+u^{\mu}$. The small deviations $u^{\mu}$ are considered as basic perturbation variables besides the metric perturbations.  They can be similarly decomposed as: $u^{\mu}=\{u^{0},\partial_{i}u+u_{i}\}$, where $u^0$, $u$ are scalar perturbations and $u_i$ is transverse and classified to be vector perturbations.

In the matter sector, the scalar field $\phi$ is decomposed as $\phi(\eta, \vec{x})=\phi(\eta)+\delta\phi$.
Other matter is assumed to have the following perturbed energy-momentum tensor:
\begin{eqnarray}
{T^{0}}_{0}&=&\bar{\rho} +\delta\rho, \nonumber\\
{T^{0}}_{i}&=&(\bar{\rho}+\bar{p}) (\partial_{i}v+v_{i}),\nonumber\\
{T^{i}}_{j}&=&(\bar{p} +\delta p)\delta^{i}_{j} + \partial_{i}\partial_{j}\sigma -(\frac{1}{3} \Delta\sigma)\delta_{ij} +\partial_{(i}\sigma_{j)}+\sigma_{ij}~.
\end{eqnarray}
where $v_{i}$, $\sigma_{i}$ are transverse vector perturbations, and $\sigma_{ij}$ is transverse and traceless tensor perturbation. In this energy-momentum tensor, $\bar{\rho}$ and $\bar{p}$ are used to denote the background energy density and pressure, viscosities are also included.

Linear perturbation equations are obtained by substituting above parametrizations to Eqs.(\ref{metric equation}) and (\ref{scalar EOM}). We will transform them to the Fourier space, so that for arbitrary function, we have $f(\eta, \vec{x}) = \frac{1}{(2\pi)^{3/2}} \int d^{3}k f(\eta, \vec{k}) e^{i\vec{k}\cdot \vec{x}}$. Even though we do not impose any gauge condition on the affine connection, we still have the freedom to do so on the metric and matter perturbations in order to simplify the calculations. Here we will choose the `conformal Newtonian gauge': $B=E=0$ and $E_{i}=0$. We would like to stress that since we use the covariant approach to keep the model diffeomorphism invariant, the conformal Newtonian gauge is always available. This is different from many other works, where at the beginning the coincident gauge ${\Gamma^{\lambda}}_{\mu\nu}=0$ has been fixed for both the background and perturbations, so there would be no displacements of $y^{\mu}$ relative to $x^{\mu}$, i.e., $u^{0}=u=0$ and $u_{i}=0$. Then one should be careful that once fixed to the coincident gauge, further gauge condition on the metric and matter perturbations may not be available. 

Since in the linear cosmological perturbation theory, the scalar, vector and tensor perturbations evolve independently, we will deal with them separately. As implied in the discussions about the background equations, both $N^{\mu\nu}$ in Eq. (\ref{metric equation}) and $Q\widetilde{Q}$ in Eq. (\ref{scalar EOM}) vanish at the background. 
We find that they also vanish at the linear order if only the scalar perturbations are considered.  This means the parity violating modification in this model has no effect on the scalar perturbations at the linear order, and the scalar perturbation equations are the same as those in GR:
\begin{align}
&\psi-A=-a^2\sigma~,\nonumber\\
&2\psi'+2\mathcal{H}A = a^2(\bar{\rho}+\bar{p})v + \phi'\delta\phi~, \nonumber\\
&2k^2\psi+6\mathcal{H}(\psi'+\mathcal{H}A)=-a^2\delta \rho + (\phi^{'2}A-\phi'\delta\phi'-a^2V_{\phi}\delta\phi)~, \nonumber\\
&2\psi'' + 2 \mathcal{H}(A'+2\psi') +2(2\mathcal{H}' + \mathcal{H}^2)A = a^2 (\delta p+\frac{2}{3}k^2\sigma) + (\phi'\delta\phi'-A{\phi^{'2}} - a^2V_{\phi}\delta\phi)~.	
\end{align} 
And we also have the same perturbed Klein-Gordon equation,
\begin{eqnarray}
\delta\phi'' + 2\mathcal{H}\delta\phi'+k^2\delta\phi-\phi'(3\psi'+A')+2a^2V_{\phi}A+a^2V_{\phi\phi}\delta\phi=0.
\end{eqnarray} 
Hence, to look for the difference between this model and GR, we should turn to consider the vector and tensor perturbations.

For vector perturbations, we find that ${N^{\mu}}_{\nu}$ only has the following non-vanishing components,
\begin{eqnarray}
{N^{0}}_{i}&=&\frac{\mathcal{M}}{a^2}\epsilon_{ijk}(\partial_{j}B_{k}+\partial_{j}\partial_{0}u_{k})~,\nonumber\\ {N^{i}}_{j}&=&\frac{-\mathcal{M}}{a^2}(\epsilon_{ikm}\partial_{j}\partial_{k}u_{m} + \epsilon_{jkm}\partial_{i}\partial_{k}u_{m} )~,
\end{eqnarray}
where the quantity $\mathcal{M}$ is defined as
\begin{equation}\label{M}
\mathcal{M}\equiv 2c(2\mathcal{H}\phi+\phi')~,
\end{equation}
its importance will be seen in later discussions. 
With these, the equations for vector perturbations are modified as 
\begin{eqnarray}
-\frac{k^2 B_{i}}{2}-i\mathcal{M}\epsilon_{ijk}k_{j}(B_{k}+\partial_{0}u_{k})&=&a^2(\bar{\rho}+\bar{p}) v_{i}~,\nonumber\\
-\mathcal{H}B_{i}-\frac{1}{2}B'_{i} +i\mathcal{M}\epsilon_{ikm}k_{k}u_{m} &=&\frac{1}{2}a^{2}\sigma_{i}~.
\end{eqnarray}
It will be more useful if we decompose the vector perturbations by the circular polarization bases $e^{L}_{i}$ and $e^{R}_{i}$:
\begin{eqnarray}	
u_{i} = u^{L}e^{L}_{i} + u^{R}e^{R}_{i}~, \quad v_{i} = v^{L}e^{L}_{i} + v^{R}e^{R}_{i}~, 
B_{i} = B^{L}e^{L}_{i} + B^{R}e^{R}_{i}~, \quad \sigma_{i} = \sigma^{L}e^{L}_{i} + \sigma^{R}e^{R}_{i}~,
\end{eqnarray}
where the bases satisfy the relation: $i\epsilon_{ijk}n_{j}e^{A}_{k}=\lambda_{A}e^{A}_{i}$, and $A=L,R$ with $\lambda_{L}=-1$, $\lambda_{R}=1$ represent the left- and right-handed polarizations respectively, $\vec{n}$ is the unit vector of $\vec{k}$. Then the equations of vector perturbations are rewritten as,
\begin{eqnarray} \label{vectorperturbations}
k^2 B^{A}+2\lambda_{A}k\mathcal{M}(B^{A}+\partial_{0}u^{A})&=&-2a^2(\bar{\rho}+\bar{p}) v^{A}~,\nonumber\\
2\mathcal{H}B^{A}+\partial_0 B^{A} -2\lambda_{A}k\mathcal{M}u^{A} &=&-a^{2}\sigma^{A}~.
\end{eqnarray}
For simplicity we drop the sources, i.e., we set $v^A=0,~\sigma^A=0$. Therefore using the first equation, we obtain a relation between $u^{A}$ and $B^{A}$,
\begin{eqnarray}\label{da}
B^{A}=D_{A}\partial_0 u^{A}~,~{\rm with}~	D_{A}=\frac{- 2\lambda_{A}\mathcal{M}}{   k + 2\lambda_{A}\mathcal{M} }.
\end{eqnarray}
then substitute it into the second equation of Eq. (\ref{vectorperturbations}), we obtain a second order differential equation for vector perturbations,
\begin{eqnarray}\label{dynamical B}
(f_A u^A)''+(k^2+2\lambda_A k \mathcal{M}-\frac{f_A''}{f_A})(f_Au^A)=0~,
\end{eqnarray}
where $f_A\equiv a\sqrt{|\mathcal{M}/(k^2+2\lambda_Ak\mathcal{M})|}$. Note that if the coupling constant vanishes, i.e., $c=0$ so that $\mathcal{M}=0$, the perturbative quantities $u^{A}$ and $B^{A}$ will decouple, and one can only obtain $B^A=0$ from Eq. (\ref{vectorperturbations}) if sources are absent. In this case, the second order equation (\ref{dynamical B}) reduces to the meaningless identity $0=0$. 

One can see that Eq. (\ref{dynamical B}) is a typical wave equation which generally describes a wave propagating in the media. We know that in GR the vector perturbations do not propagate, they represent some constraints rather than dynamical modes. In the model considered here, however, Eq. (\ref{dynamical B}) shows unambiguously that the vector perturbations are indeed dynamical degrees of freedom. This is the first new result we obtained for the model (\ref{action full}). We note that the feature of promoting the vector perturbations to the dynamical degrees of freedom also appeared in some other modified gravity models, for instance the gravity model with a Weyl term \cite{Deruelle:2010kf}. As dynamical components, the vector perturbations can be generated at the early universe (e.g., the epoch of inflation) from vacuum fluctuations. Then at later times, the vector perturbations would leave important effects on the large scale structure and CMB anisotropies, these might be confirmed or excluded by the cosmological probes. 

Due to the parity violation, the left- and right-handed polarized components of the vector perturbations have different propagating velocities and their amplitudes damped with different rates. 
One can read from Eq. (\ref{dynamical B}) that the dispersion relations $\omega_A^2=k^2+2\lambda_A k \mathcal{M}$ is handedness dependent. This leads to different propagating velocities. In terms of the terminology of Ref. \cite{Zhao:2019xmm}, this phenomenon is called ``velocity birefringence".  In addition, $f_A$ in Eq. (\ref{dynamical B}) is also handedness dependent, this means the amplitudes of $u^A$ with $A=L, R$ damped with different rates 
as they propagate in the universe. This phenomenon is dubbed ``amplitude birefringence" and happened for the GWs in the CS modified gravity. So the model (\ref{action full}) presented both velocity and amplitude birefringence phenomena in the vector perturbations. 

Now we turn to the properties of tensor perturbations, these have been studied in Ref. \cite{Conroy:2019ibo} and in other papers. According to the decomposition (\ref{decomposition}), the symmetric tensor perturbations $h_{ij}$ are transverse and traceless. So they only have two independent components, these correspond to two polarizations of GWs. 
When consider the parity violating extension, we find that only the $(i, j)$ components of ${N^{\mu}}_{\nu}$ do not vanish,
\begin{eqnarray}
{N^{i}}_{j} = \frac{\mathcal{M}}{a^2} (\partial_{m}h_{ki}\epsilon_{jmk} + \partial_{m}h_{kj}\epsilon_{imk})~,
\end{eqnarray}
so the equation for tensor perturbations is 
\begin{align}
h_{ij}''+2\mathcal{H}h_{ij}'+k^2 h_{ij}+2i \mathcal{M}k_{m}(h_{ki}\epsilon_{jmk}+h_{kj}\epsilon_{imk})=-2a^{2}\sigma_{ij}~.
\end{align}
Next, just like we have done for the vector perturbations, we also expand the tensor perturbations in terms of the circular polarization bases $e^{L}_{ij}$ and $e^{R}_{ij}$,
\begin{eqnarray}
h_{ij}=h^{L}e^{L}_{ij}+h^{R}e^{R}_{ij}, \quad \sigma_{ij}=\sigma^{L}e^{L}_{ij}+\sigma^{R}e^{R}_{ij}~.
\end{eqnarray}
The bases satisfy the relation: $\epsilon_{ilk}n_{l}e^{A}_{jk}=i\lambda_{A}e^{A}_{ij}$. Again $A=L,R$ and $\lambda_{L}=-1$, $\lambda_{R}=1$. 
With these circular polarization bases, the above equation can be rewritten as, 
\begin{eqnarray}
{h^{A}}''+2\mathcal{H}{h^{A}}'+k^{2}h^{A}-4\lambda_{A}\mathcal{M}k h^{A}=-2a^2\sigma^{A}~.
\end{eqnarray}
This equation indicates that the parity violating interaction $\phi(x)Q\widetilde{Q}$ neither changes the number of independent polarization modes of GWs nor causes mixing between two circularly polarized modes. Its salient effect on GWs is inducing handedness dependent dispersion relation $\omega_{A}^{2}=k^{2}-4\lambda_{A}\mathcal{M}k$ and making a difference between the propagating velocities of the two helicities of GWs . This is the velocity birefringence phenomenon in GWs and signals again the parity violation of the model. Considering small coupling constant $c$ and slow evolution of $\phi$,  one can find from the dispersion relation that GWs with different helicities have different phase velocities
\begin{eqnarray}\label{phase velocity}
v^{A}_{p}=\frac{\omega_{A}}{k} \approx 1-\frac{2\lambda_{A}\mathcal{M}}{k}~,
\end{eqnarray}
and the same group velocity up to the order $\mathcal{O}$($\mathcal{M}^2$)~,
\begin{eqnarray} \label{group velocity}
v_{g}^{A}=\frac{d\omega_{A}}{dk} \approx 1+\frac{2\mathcal{M}^{2}}{k^2}~.
\end{eqnarray}
This property of tensor perturbations is the same as that of NYTG model \cite{Li:2020xjt,Li:2021wij}. Furthermore, this is an infrared effect, because it is important only at low frequencies (large length scales). On the contrary, in the CS modified gravity \cite{Alexander:2009tp} the modification by the CS term gives rise to the amplitude birefringence in GWs, and as shown in Ref.  \cite{Dyda:2012rj} it is an ultraviolet effect because it becomes important at high momentum scales.  

The propagating velocities of GWs in Eq. (\ref{phase velocity}) are different from the speed of light which is set to one. This deviation is tightly constrained by current GWs experiments. The coincident detections of GW170817/GRB170817A \cite{LIGOScientific:2017vwq, LIGOScientific:2017zic} impose tight bounds on the propagating speed of GWs: $-3\times10^{-15}<v_{p}^{A}-1<+7\times 10^{-16}$. In the case discussed here, this means $|\mathcal{M}|/k<3.5\times 10^{-16}$.  Considering the frequency $k/a\sim 100 $Hz of LIGO, this in turn constrain the quantity $\mathcal{M}$ as $|\mathcal{M}|/a<2.3\times 10^{-38}$GeV. 
In addition, the velocity birefringence in this model brings modifications to the GW phase. Confronting such modifications with data of GWs events of binary black hole merges observed by LIGO-Virgo, more stringent constraint on $\mathcal{M}$ has been recently imposed in Ref. \cite{Wu:2021ndf}: $|\mathcal{M}|/a<1.6\times 10^{-42}$GeV. Please note that the quantity $\mathcal{M}$ relates the energy scale parameter $M_{PV}$ of Ref. \cite{Wu:2021ndf} as $-4\mathcal{M}/a=M_{PV}$. 

The dispersion relation $\omega_{A}^{2}=k^{2}-4\lambda_{A}\mathcal{M}k$ in this model indicates another instability at very large scales. When $k<4|\mathcal{M}|$, $\omega_{A}^{2}$ has the possibility to be negative and one of circularly polarized tensor modes increase with time. Because $|\mathcal{M}|$ is constrained tightly by current GWs experiments, this instability happens at large length scales $k/a<6.4\times 10^{-42}$GeV. This bound corresponds to current horizon scale. The growth of perturbation at super horizon scales is common for systems containing gravity. This instability is different from the ghost instability. It is not observable for the experiments within the horizon and will not cause severe problems when quantizing the perturbation theory. 

Here, we would like to mention some observational signatures of parity violating gravity models. These models present two major effects in GWs, i.e., the amplitude birefringence and velocity birefringence. The amplitude birefringence can modify the GW amplitudes of two circularly polarized modes, and the velocity birefringence can modify the GW phases of these two modes. These two effects can be quantified by the specific parametrized post-Einsteinian (PPE) parameters \cite{Zhao:2019xmm} and the PPE parameters can be tested or constrained by GWs data \cite{Wang:2020cub}. The expressions of PPE parameters are usually different for different models, which can be easily distinguished in data analysis \cite{Yunes:2009ke,Cornish:2011ys}. In addition, there are waveform independent methods to test the parity violation of gravity by GWs observations. For instances, for amplitude birefringence phenomenon, one can make statistically analysis on the ratio of two circular polarization modes by analyzing the distribution of inclination angles of a sample of GW events \cite{Okounkova:2021xjv}, for velocity birefringence phenomenon, one can decompose the left- and right-handed modes and directly compare the difference of arriving times \cite{Zhao:2019szi}. For the scenarios without parity violation, the birefringence phenomenon is absent, such as the $f(T)$ model \cite{Li:2018ixg}. Among the parity violating models, the amplitude birefringence and the velocity birefringence can be easily distinguishes. However, degeneracies may exist in some cases. For example, both the model studied in this paper and the NYTG model \cite{Li:2020xjt} predicted velocity birefringence but no amplitude birefringence in GWs. This degeneracy is broken by different predictions in the vector perturbations. The model studied in this paper predicted non-trivial modifications to the vector perturbations, but the NYTG model brings no modification to the vector perturbations when compared with GR. 
As discussed before, all the observational signals of the model studied in this paper depends on the quantity $\mathcal{M}$ defined in Eq. (\ref{M}). It is proportional to the coupling constant $c$. The experiments are sensitive to $\mathcal{M}$ and currently put stringent constraints on it as mentioned before. Obtaining direct constraint or measurement on $c$ from data depends on the specific model of the scalar field $\phi$, this is beyond the scope of this paper. 

\subsection{Quadratic Actions for Scalar, Vector and Tensor Perturbations}

For a full analysis on the linear cosmological perturbations, having the perturbation equations is not enough. We should also care about the actions which give the perturbation equations through the variational principle. 
Furthermore, the actions are indispensable when applying the gravity model to the early universe, such as the inflationary universe, where in order to produce the primordial perturbations we need the actions to quantize the perturbations. For linear perturbations, we need the quadratic actions. Now we will calculate the quadratic actions for the scalar, vector and tensor perturbations respectively from the original action (\ref{action full}), and the background equations (\ref{beom}) are assumed to be already known. For a further simplification, we ignore the matter other than the scalar field, that is $S_{m}=0$. This is reasonable because the quadratic actions are usually applied for the early universe, at that time the universe was dominated by a scalar field (such as the inflaton) which may be identified with $\phi$ of this model, and other matter are negligible. 

Again, we will choose the conformal Newtonian gauge to simplify our calculations. The affine connection vanishes at the background, but in the perturbed universe it is determined by the displacements $u^{\lambda}$, 
\begin{eqnarray} \label{quadratic gamma}
{\Gamma^{\lambda}}_{\mu\nu}=\partial_{\mu}\partial_{\nu}u^{\lambda} - \partial_{\mu}\partial_{\nu}u^{\beta}\partial_{\beta}u^{\lambda}~.
\end{eqnarray}
As mentioned before, the displacements are decomposed into the scalar and vector perturbations: $u^{\mu}=\{u^{0},\partial_{i}u+u_{i}\}$.
For the scalar perturbations, one can find that no matter which gauge we choose, the parity violating term $\phi Q\widetilde{Q}$ in the action (\ref{action full}) is always zero up to the second order.  This is consistent with the scalar perturbation equations in the previous subsection.  The parity violating term doesn't influence the scalar perturbations and the final quadratic action for the scalar perturbations is the same as that in GR with a minimally coupled scalar field. It is very convenient to introduce a gauge invariant variable $\zeta=-(\psi+\mathcal{H}\delta\phi/\phi')$, which denotes the curvature perturbation of the hypersurfaces with constant $\phi$ field. Using this gauge invariant variable $\zeta$ and eliminating the constraints, the quadratic action for the scalar perturbations is finally written as
\begin{eqnarray}
S_S^{(2)}=\int d^{4}x \; z^2  \left( \zeta^{'2} - \partial_{i}\zeta\partial_{i}\zeta \right)~,
\end{eqnarray}
where $z^2=a^2\phi^{'2}/(2\mathcal{H}^2)$.

For the vector perturbations, the conformal Newtonian gauge is $E_{i}=0$. With it, the quadratic action has the initial form, 
\begin{align} \label{quardratic action vector}
 S_V^{(2)}=\int d^{4}x \,  a^{2} \{  \frac{1}{4} \partial_{j}B_{i}\partial_{j}B_{i} 
-\frac{\mathcal{M}}{2}\epsilon_{ijk} (B_{i}\partial_{j}B_{k} +2B_{i}\partial_{j}u'_{k} +u'_{i}\partial_{j}u'_{k} - \partial_{m}u_{i}\partial_{j}\partial_{m}u_{k}) \}~. 
\end{align}
One can see that the variables $B_{i}$ are not dynamical fields, the variation of the action (\ref{quardratic action vector}) with respect to them gives the constraint equation,
\begin{eqnarray}\label{constrant B}
-\Delta B_{i}  +2\mathcal{M}\epsilon_{ijk} \left( \partial_{j}B_{k} +\partial_{j}u'_{k} \right)=0~.
\end{eqnarray}
We need to solve this constraint equation to get $B_{i}$ as a function of $u_{i}$, and then substitute this relation back into the action (\ref{quardratic action vector}) to eliminate the constraints.  It is easy to do so if  we expand the constraint in Fourier space by the circular polarization bases, 
\begin{eqnarray}
B_{i}(\eta, \vec{x}) = \sum_{A=L,R} \int \frac{d^3k}{(2\pi)^{3/2}} B^{A}(\eta,\vec{k}) e^{A}_{i}(\vec{k}) e^{i\vec{k}\cdot \vec{x}}~,
\end{eqnarray}
where the base vectors are orthogonal: $e^{A}_{i}e^{B*}_{i}=\delta_{AB}$. Then from the constraint equation (\ref{constrant B}), we again obtain the relation between $u^{A}$ and $B^{A}$, i.e., Eq. (\ref{da}). 
With these, the quadratic action for the vector perturbations (\ref{quardratic action vector}) can be finally written as 
\begin{eqnarray} \label{quadratic action vector newton}
S_V^{(2)}	=\frac{1}{2}\sum_{A=L,R} \int d\eta d^3k \, z_A^2\left(  {u^{A}}'{u^{A*}}'  -\omega^{2}_{A} {u^{A}}{u^{A*}}\right)~,
\end{eqnarray}
where $\omega^{2}_{A}= k^2 + 2\lambda_{A}\mathcal{M}k$, and 
\begin{eqnarray}\label{z2}
z_A^2=\frac{\lambda_{A}a^2\mathcal{M}k^2}{k+2\lambda_{A}\mathcal{M}}.
\end{eqnarray}
It can be seen first that the vector perturbations are truly dynamical, the two modes propagate in the background spacetime with different velocities, i.e., the phenomenon of velocity birefringence, because $\omega_A^2$ in this quadratic action is indeed handedness dependent. Second, the factor $z_A^2$ is an overall factor of this quadratic action, it is also handedness dependent and leads to amplitude birefringence phenomenon. Third, the sign of $z_A^2$ determines whether the vector modes are ghosts or not. At low momentum scales (large length scales), where $k\ll |2\lambda_{A}\mathcal{M}|$ ,~$z_A^2 \backsimeq a^2k^2/2$, and it is positive definite. So both the two components of the vector perturbations have the right sign in their kinetic terms at these scales. But at high momentum scales (small length scales), where $k\gg |2\lambda_{A}\mathcal{M}|$, and $z_A^2 \backsimeq \lambda_{A}a^2\mathcal{M}k$. Since $\lambda_{A}=\pm1$, it means that one component of the vector perturbations must be a ghost mode because its kinetic term has the wrong sign. The existence of ghost mode will make the Hamiltonian not bounded from below and in the quantum theory of the perturbations the ghost mode cause significant problems such as the vacuum instability: the vacuum will decay into particles in a very fast rate. In another word, this model suffers from the pathologies brought by the ghost mode. This is the second new result we obtained for the model 
(\ref{action full}).

The quadratic action for the tensor perturbations can be obtained in a straightforward way, 
\begin{align}
S_T^{(2)}=\int d^{4}x a^2 [ \frac{1}{8}(h_{ij}'h_{ij}'- \partial_{l}h_{ij}\partial_{l}h_{ij})+\frac{\mathcal{M}}{2}\epsilon_{ijk}h_{im}\partial_{j}h_{km}]~.
\end{align}
In terms of the circular polarization bases with $e^{A}_{ij}e^{B*}_{ij}=\delta_{AB}$, the quadratic action can be finally written as
\begin{align}
S_T^{(2)}=\sum_{A=L,R} &\int d\eta d^3k \frac{a^2}{8} [ h^{A'}h^{A*'} -(k^2-4\lambda_{A}\mathcal{M}k)h^{A}h^{A*} ].
\end{align}
From it, one can see that both components of GWs are healthy. The variation of this action with respect to $h^A$ lead to the equations of tensor perturbations discussed in the previous subsection.

\subsection{Further Discussions on the Quadratic Action of Vector Perturbations}\label{discussion}

In this subsection, we will recalculate the quadratic action for the vector perturbations in a different way. We return to the conventional operation used extensively in studies of STG theory: taking the coincident gauge so that the affine connection ${\Gamma^{\lambda}}_{\mu\nu}=0$ once and for all. This means the functions $y^{\mu}$ always match with $x^{\mu}$ even in the perturbed universe, and the vector perturbations $u_i$ or $u^A$ mentioned before are absent here. Under this gauge, we will not take any other gauge condition on the metric and matter sector because extra gauge conditions may not be compatible with the coincident gauge. 
With these considerations, the quadratic action for the vector perturbations is
\begin{align} \label{quadratic action vector coincident}
S_V^{(2)}=\int d^{4}x \,  a^{2} [ \frac{1}{4}\left(\partial_{j}B_{i}\partial_{j}B_{i} +2\partial_{j}B_{i}\partial_{j}E'_{i} +\partial_{j}E'_{i}\partial_{j}E'_{i}\right)
-\frac{\mathcal{M}}{2}\epsilon_{ijk}\left( B_{i}\partial_{j}B_{k} - \partial_{m}E_{i}\partial_{j}\partial_{m}E_{k}\right) ]~.
\end{align}
Again, the variables $B_{i}$ are not dynamical fields, in stead they represent constraints and satisfy the equation,
\begin{eqnarray}
-\Delta B_{i} -\Delta E'_{i} +2\mathcal{M}\epsilon_{ijk}\partial_{j}B_{k}=0~.
\end{eqnarray}
After expanding the constraints in Fourier space by the circular polarization bases, we obtain the relation $B^{A}=-(1+D_{A}){E^{A}}'$. Then substitute this relation back into the action (\ref{quadratic action vector coincident}), the quadratic action can be rewritten as 
\begin{eqnarray}\label{quadratic action vector coincident final}
S_V^{(2)}	=\frac{1}{2}\sum_{A=L,R} \int d\eta d^3k \, z_A^2\left(  {E^{A}}'{E^{A*}}'  -\omega^{2}_{A} {E^{A}}{E^{A*}}\right)~.
\end{eqnarray}
During above calculations, the notations $D_A$, $\omega_A^2$ and $z_A^2$ are the same with those defined in previous subsection. The quadratic action (\ref{quadratic action vector coincident final}) is the same as (\ref{quadratic action vector newton}) obtained in previous subsection except the variables used to represent the vector perturbations are different. This depends on the gauge choice.  

For vector perturbations, we can define gauge invariant variables $V_{i}=E_{i}-u_{i}$. Thus in the conformal Newtonian gauge $V_{i}=-u_{i}$, and in the coincident gauge $V_{i}=E_{i}$. Then we can find that Eq.(\ref{quadratic action vector newton}) and Eq.(\ref{quadratic action vector coincident final}) are totally the same if we rewrite both of them in terms of the gauge invariant variables $V_{i}$. 

\section{Summary} \label{summary}

Possible parity violations in the gravity sector have attracted lots of interests in recent years. In this paper, we studied a simple parity violating gravity model which modifies STGR by a coupling $\phi(x)Q\widetilde{Q}$ between a scalar field and the non-metricity tensor. This coupling was known to cause the velocity birefringence phenomenon in GWs \cite{Conroy:2019ibo}. In this paper we applied this model to cosmology and investigated in detail its effects on the background evolution of the universe and attached more importance to the cosmological perturbations. We found that besides the velocity birefringence in the tensor perturbations the coupling $\phi(x)Q\widetilde{Q}$ does not affect the background evolution and the scalar perturbations. But it has important modifications to the vector perturbations. Due to this coupling, the vector perturbations are promoted to be dynamical degrees of freedom, this is different from GR where the vector perturbations are just constraints. More importantly, we found that one component of the vector perturbations must be a ghost, this leads to the significant difficulty of vacuum instability in the quantum theory of perturbations. Even though the ghost mode was found in the cosmological perturbation theory, we believe it is originated from the defect of the model itself. The coupling $\phi(x)Q\widetilde{Q}$ can be applied on top of an $f(\mathbb{Q})$ modification, e.g., the one leads to successful confrontation with the data \cite{Anagnostopoulos:2021ydo}. This further modification by replacing $\mathbb{Q}$ in the model (\ref{action full}) with 
$f(\mathbb{Q})$ cannot help to solve the ghost problem in the vector perturbations. We know from the quadratic action (\ref{quadratic action vector newton}) that the ghost mode originates from handedness dependence of the overall factor $z_A^2$, as shown in Eq. (\ref{z2}). This dependence also leads to the amplitude birefringence phenomenon. It is a feature of parity violation. Any further parity conserving modification, such as $f(\mathbb{Q})$, cannot eliminate the handedness dependence of $z_A^2$, so is not able to solve the ghost problem. However, taking into account more parity violating extensions is expected to eliminate the ghost mode.

It is interesting to make comparisons among three simple parity violating gravity models. Each slightly modifies GR only by a parity violating interaction in the Lagrangian. The first one is the CS modified gravity \cite{Jackiw:2003pm,Alexander:2009tp} which modifies GR by a CS term $\phi(x) R\widetilde{R}$ within the framework of Riemannian geometry. This model produces the amplitude birefringence of GWs, but is pathological due to the ghost mode in the tensor perturbations. The second one is the model (\ref{action full}) considered in this paper. It modifies GR by the coupling $\phi(x) Q\widetilde{Q}$ within the framework of STG theory. As we have investigated in this paper, this model generates velocity birefringence phenomenon in GWs and both velocity and amplitude birefringence phenomena in vector perturbations. It also fails to be a feasible model due to the ghost mode in the vector perturbations. The third one is the NYTG model  \cite{Li:2020xjt,Li:2021wij} which modifies GR by a NY term $\phi(x)T\widetilde{T}$ within the framework of TG theory. This model produces the velocity birefringence phenomenon in GWs. Detailed investigations on the scalar, vector and tensor perturbations \cite{Li:2020xjt,Li:2021wij} showed that this model is free from the ghost mode. Hence compared to the former two models, the third one is more successful.

\subsection*{Acknowledgements}
We thank Wen Zhao for helpful discussions. This work is supported by NSFC under Grant Nos. 12075231, 11653002, 12047502 and 11947301.

\bibliographystyle{utphys}
\providecommand{\href}[2]{#2}\begingroup\raggedright
\endgroup

\end{document}